\documentclass[12pt]{article}
\usepackage{graphics}
\usepackage{amssymb}

\begin{document}

\title{Persistent currents due to point obstacles}
\author{T.~Cheon$^{a}$ and P.~Exner$^{b,c}$}
\date{}
\maketitle
\begin{quote}
{\small \em a) Laboratory of Physics, Kochi University of Technology, 
\\ \phantom{e)x}Tosa Yamada, Kochi 782-8502, Japan
\\ b) Department of Theoretical Physics, Nuclear
Physics Institute, \\ \phantom{e)x}Academy of Sciences, 25068 \v
Re\v z, Czech Republic \\
 c) Doppler Institute, Czech Technical University, B\v{r}ehov{\'a} 7,\\
\phantom{e)x}11519 Prague, Czech Republic \\
 \rm \phantom{e)x}taksu.cheon@kochi-tech.ac.jp, exner@ujf.cas.cz}
\vspace{8mm}

\noindent {\small We discuss properties of the two-dimensional
Landau Hamiltonian perturbed by a family of identical $\delta$
potentials arranged equidistantly along a closed loop. It is
demonstrated that for the loop size exceeding the effective size
of the point obstacles and the cyclotronic radius such a system
exhibits persistent currents at the bottom of the spectrum. We
also show that the effect is sensitive to a small disorder.}
\end{quote}


\section{Introduction}

Persistent currents in rings threaded by a magnetic flux are one
of the characteristic features of mesoscopic systems -- see, e.g.,
\cite{CGR, CWB} and numerous other theoretical and experimental
papers where they were discussed. Recall that for a charged
particle (an electron) confined to a loop $C$ the effect is
contained in the dependence of the corresponding eigenvalues $E_j$
on the flux $\phi$ threading the loop measured in the units of
flux quanta, $2\pi\hbar c|e|^{-1}$. Specifically the derivative
$\partial E_j/\partial \phi$ equals $-{1\over c}I_j$, where $I_j$
is the persistent current in the $j$--th state. In particular, if
the particle motion on the loop is free, we have
 \begin{equation} \label{ideal}
E_j(\phi) = {\hbar^2\over 2m^*} \left( 2\pi\over L\right)^2
(j+\phi)^2,
 \end{equation}
where $L$ is the loop circumference, so the currents depend
linearly on the applied field.

The above example represents an ideal case when the particle is
strictly confined to the loop. On the other hand, there are many
situations when a particle in a magnetic field can be transported
being localized essentially in the vicinity of a barrier or an
interface. A prominent example are the edge currents \cite{Ha,
MS}. They attracted a new wave of interest recently when in
several papers \cite{BP, FGW, FM, MMP} conditions were derived
under which such a transport remained preserved in the presence of
a disorder.

Even weaker interaction giving rise to a magnetic transport, this
time without a classical analogue, was found in \cite{EJK}, where
the Landau Hamiltonian was perturbed by a straight equidistant
array of $\delta$ potentials. It was shown that such an operator
has bands of absolutely continuous spectrum in the gaps between
the Landau levels. An independent evidence for the transport in
this situation was found in \cite{Ue}.

The question we are going to ask in this paper is the following.
If we take a finite array of identical $\delta$ potentials in the
plane making a loop, does the mentioned transport give rise to
persistent currents? As we will see the answer is in general
positive if the effective radii of the $\delta$ potentials
(specified below -- see (\ref{effrad}))  and the cyclotronic
radius corresponding to the field intensity are much smaller than
the loop size. We also discuss the effect of disorder manifested
by randomization of the coupling constants and show that the
effect is sensitive to such perturbations: in an example with a
ring of 12 obstacles a variation larger than 1\% leads to a
localization.


\setcounter{equation}{0}
\section{The model}

\subsection{2D magnetic systems with point obstacles}

As we said above, we are going to study a charged particle in the
plane subject to a homogeneous magnetic field perpendicular to the
plane interacting with an equidistant array of point obstacles
placed at points $\vec a_j,\: j=1,\dots,N$, belonging to a closed
loop. The simplest example is the situation when the latter is a
circle,
 \begin{equation} \label{circle}
\vec a_j= \left( R\,\cos{2\pi j\over N},\, R\,\sin{2\pi j\over N}
\right)\,,\; j=1,\dots,N\,.
\end{equation}
For the sake of brevity we shall write $\vec a:= (\vec a_1, \dots,
\vec a_N)$. The obstacles will be modeled by $\delta$ potentials
so formally the Hamiltonian of the system can be written as
 \begin{equation} \label{Hamiltonian}
H = \left(-i\vec\nabla -\vec A(\vec x)\right)^2 +\sum_{j=1}^N
\tilde\alpha_j \delta(\vec x\!-\!\vec a_j)\,,
 \end{equation}
where $\vec A$ is the appropriate vector potential which we choose
here in the circular gauge,
 $$ 
\vec A(\vec x) = \left( -{1\over 2}By,\, {1\over 2}Bx \right)\,.
 $$ 
To make things simple we have used rationalized units in
(\ref{Hamiltonian}) putting $\hbar =c =e =2m^* =1$. On the other
hand, the two-dimensional $\delta$ interaction is an involved
object. To give it meaning and to specify the true coupling
parameter which would replace the formal $\tilde\alpha_j$'s, we
use the definition based on self-adjoint extensions which is
discussed thoroughly in \cite{AGHH}, see also \cite{DO, GHS} for
the situation with a magnetic field. A $\delta$ interaction is
determined by means of the boundary conditions
\begin{equation} \label{bc}
L_1(\psi,\vec a_j)+ 2\pi\alpha_j L_0(\psi,\vec a_j)=0\,, \quad
j=1,\dots,N\,,
\end{equation}
which couple the generalized boundary values
 $$ 
L_0(\psi,\vec a) := \lim_{|\vec x-\vec a|\to 0}\, {\psi(\vec
x)\over \ln |\vec x\!-\!\vec a|}\,, \; L_1(\psi,\vec a) :=
\lim_{|\vec x-\vec a|\to 0} \Bigl\lbrack \psi(\vec x)\!-\!
L_0(\psi,\vec a)\, \ln |\vec x\!-\!\vec a| \Bigr\rbrack\,.
 $$ 
We now put $\alpha:= (\alpha_1, \dots, \alpha_N)$ and denote by
$H^B_{\alpha,\vec a}$ the Hamiltonian of our model, which acts as
free outside the interaction support,
 $$
H^B_{\alpha,\vec a}\psi(\vec x) = \left(-i\vec\nabla -\vec A(\vec
x)\right)^2 \psi(\vec x)
 $$
for $\vec x\not\in \vec a$, and its domain consists of all
functions which belong to the Sobolev class $W^{2,2}(\mathbb{R}^2
\setminus\vec a)$ and satisfy the boundary conditions (\ref{bc}).

The parameters $\alpha_j$ characterizing the ``strength'' of the
interaction are not trivial \cite{ES,CS}; this is clear from the fact that the
free (Landau) Hamiltonian, which we denote as $H^B$, corresponds
to $\alpha_j=\infty,\: j=1,\dots,N$. The difference between
$\alpha_j$ and the formal coupling constants used in
(\ref{Hamiltonian}) reflects the nontrivial way in which the
two-dimensional point interaction arises in the limit of scaled
potentials. To understand the $\alpha_j$'s it is more natural to
associate them with the scattering length: by \cite{ES} the point
interaction with the parameter $\alpha$ reproduces the low-energy
behaviour of the scattering upon an obstacle of the radius $\rho$
provided
\begin{equation} \label{effrad}
\alpha= {1\over 2\pi}\, \ln\rho\,.
\end{equation}

\subsection{Spectral properties of $H^B_{\alpha,\vec a}$}

The spectrum of the free operator $H^B$ is the series of
infinitely degenerate eigenvalues $B(2n+1),\: n=0,1,\dots\,$. A
finite number of point interactions leaves the essential spectrum
intact so $\sigma_\mathrm{ess}(H^B_{\alpha,\vec a})$ consists
again of the Landau levels. Moreover, it follows from general
principles \cite[Sec~.8.3]{We} that $N$ point interactions can
give rise to at most $N$ eigenvalues in each gap of $\sigma(H^B)$.
It is this discrete spectrum which is the object of our interest.

To find it we can make use of Krein's formula \cite[App.~A]{AGHH}
which expresses the resolvent kernel of $H^B_{\alpha,\vec a}$ by
means of that of the free operator, specifically
 \begin{equation} \label{krein}
(H^B_{\alpha,\vec a}-\!z)^{-1}(\vec x,\vec x')= G_B(\vec x,\vec
x';z) + \sum_{j,m=1}^N [\Lambda^B_{\alpha,\vec a}(z)]^{-1}_{jm}
G_B(\vec x,\vec a_j;z) G_B(\vec a_m,\vec x';z),
 \end{equation}
where $\Lambda^B\equiv \Lambda^B_{\alpha,\vec a}(z)$ is the
$N\times N$ matrix with the elements
   $$
\Lambda^B_{jm}: = \left(\alpha_j-\xi(B;z)\right) \delta_{jm}
-G_B(\vec a_j,\vec a_m;z) (1\!-\!\delta_{jm})\,,
   $$
where $G_B(\vec x,\vec x';z)$ is the resolvent kernel of $H^B$ and
$\xi(B;z)$ its regularized value at the singularity,
 \begin{equation} \label{xi}
\xi(B;z) := \lim_{\vec x'\to \vec x} \left\lbrack\, G_B(\vec
x,\vec x';z) + {1\over 2\pi}\, \ln|\vec x-\vec x'| \right\rbrack
 \end{equation}
which is independent of $\vec x$. The sought discrete spectrum is
given by the singularities of the second term at the r.h.s. of
(\ref{krein}), or in other words, by solutions of the spectral
condition
 \begin{equation} \label{spectral condition}
\det \Lambda^B_{\alpha,\vec a}(z) = 0\,.
 \end{equation}
If $z=z_0$ satisfies this, it is an eigenvalue of the operator
$H^B_{\alpha,\vec a}$ and the equation $\Lambda^B_{\alpha,\vec
a}(z_0)d=0$ has a nontrivial solution in $\mathbb{C}^N$; the
corresponding non-normalized eigenfunction is then
 \begin{equation} \label{ef}
\psi^{B,z_0}_{\alpha,\vec a} = \sum_{j=1}^N d_j G_B(\vec
a_j,\cdot\,;z_0)
 \end{equation}
as can be checked by a standard argument \cite{AGHH}.

To make use of the above formulae we need the explicit form of the
free Green's function. It is known \cite{DMM} to be
 \begin{equation} \label{freeG}
G_B(\vec x,\vec x';z) = {1\over 4\pi}\, \Phi_B(\vec x,\vec x')\,
\Gamma\left( |B|\!-\!z\over 2|B| \right)\, U\left( {|B|\!-\!z\over
2|B|},\, 1; {|B|\over 2}|\vec x\!-\!\vec x'|^2 \right)\,,
 \end{equation}
where
   $$
\Phi_B(\vec x,\vec x') := \exp\left\lbrack -{iB\over
2}(x_1\!-\!x'_1) (x_2\!+\!x'_2) - {|B|\over 4} |\vec x\!-\!\vec
x'|^2 \right\rbrack
   $$
and $U(u,v;\zeta)$ is the singular confluent hypergeometric (or
Kummer) function \cite[Chap.~13]{AS}. Using the asymptotic formula
   $$
U(u,1;\zeta) = - {1\over \Gamma(u)}\, \left\lbrack\, \ln\zeta +
\psi(u) -2\psi(1) + \mathcal(O)(\zeta^2\ln^2\zeta) \,\right\rbrack
   $$
with $\psi(u)$ being the digamma function and $-\psi(1)= \gamma=
0.577\dots$ the Euler number we find
 \begin{equation} \label{xiexplicit}
\xi(B;z) = - {1\over 4\pi}\, \left\lbrack\, \psi\left(
|B|\!-\!z\over 2|B| \right) + 2\gamma + \ln {|B|\over 2}
\,\right\rbrack\,.
 \end{equation}
In this way we have derived the explicit form of the matrix
$\Lambda^B_{\alpha,\vec a}(z)$ appearing in the condition
(\ref{spectral condition}).


\setcounter{equation}{0}
\section{The results}

In Fig.~1, we show the energy spectra of the system as a function
of the common coupling strength $\alpha_i = \alpha$. In this
example, the $Np = 12$ of $\delta$-interactions are placed
equidistantly on the ring of circumference $L = 2\pi$. We can make
this assumption without loss of generality. It is true that
scaling systems with 2D point interactions reveals peculiar
properties \cite{CS2}, but in effect changing the size of the
system is equivalent to the shift of $\alpha$ on a constant
proportional to logarithm of the scaling factor. One can see from
this figure that, for our choice of the system size, the coupling
becomes strong as the parameter approaches the value $\alpha =
-1$, in which case the eigenvalues split from the Landau levels of
unperturbed system form a bunch of negative-energy states; recall
that by(\ref{effrad}) it corresponds to the scattering length
$\rho= e^{-2\pi}$.

The properties of these states are made clear by plotting their
energies as the function of the magnetic field $B$.  We set the
strength parameter to be $\alpha = -1$. In the top lines of
Fig.~2, several lowest states are shown for for three cases $Np =
6$, $Np = 12$ and $Np = 24$. As the number of the
$\delta$-scatterers increases, one clearly sees the emergence of
parabolic $B$ dependence, which is the hallmark of the persistent
current we have been looking for. The situation becomes clearer
when we plot derivatives $dE/dB$ and $d^2E/dB^2$ as the function
of $B$.
A direct evidence of the existence of persistent current in our
system is shown in Fig.~3, in which the probability flux current
is depicted for $Np = 12$ case for the magnetic field strengths $B
= 0.5$, $B = 1.0$, $B = 1.5$ and $B = 2.0$.

As with the ``usual'' persistent currents, one has to take into
account the imperfection when thinking about the use of the effect
in designing various devices. Thus the question of the robustness
of the persistent current in our system is of interest. We test it
by allowing random fluctuation of the coupling strengths of the
$\delta$-scatterers. In Fig.~4, we plot the $E$, $dE/dB$ and
$d^2E/dB^2$ as the function of $B$ as before, but letting the
coupling strength of each scatterer to vary randomly by the amount
$\Delta \alpha$ uniformly around the central value $\alpha = -1$.
The fluctuations are taken to be $\Delta \alpha = 0.01 $, $\Delta
\alpha = 0.02$ and $\Delta \alpha = 0.03$ from the left to right.
The number of $\delta$-scatterers is again set to be $Np = 12$.
Some examples of probability current flow patterns at $B = 0.5$
for each case are shown in Fig.~5. While a more detailed analysis
of such random perturbations is required, Figs.~4-5 suggest that
the persistent current generated by the array of
$\delta$-scatterers is sensitive to small fluctuations of the
strength; one typically needs the accuracy of less the 1\% to have
the persistent current intact down to the ground state.

In conclusion, we have demonstrated a purely quantum mechanism
able to create persistent currents in systems with arrays of tiny
obstacles in magnetic field; it is the the wave nature of
electrons which is responsible for this counterintuitive
phenomenon. The result might eventually have a practical impact on
the design of the quantum ring devices, however, the effect of
localization by random perturbations deserves a deeper study.

\subsection*{Acknowledments}

P.E. appreciates the hospitality extended to him at the Kochi
University of Technology where a part of this work was done. The
research has been partially supported by GAAS and the Czech
Ministry of Education within the projects A1048101 and ME170, and
also by Grant-in-Aid for Scientific Research (C) (No. 13640413) by
the Japanese Ministry of Science and Education.
%

%
\newpage
\begin{figure}
\centerline{
\scalebox{0.60}{
\includegraphics{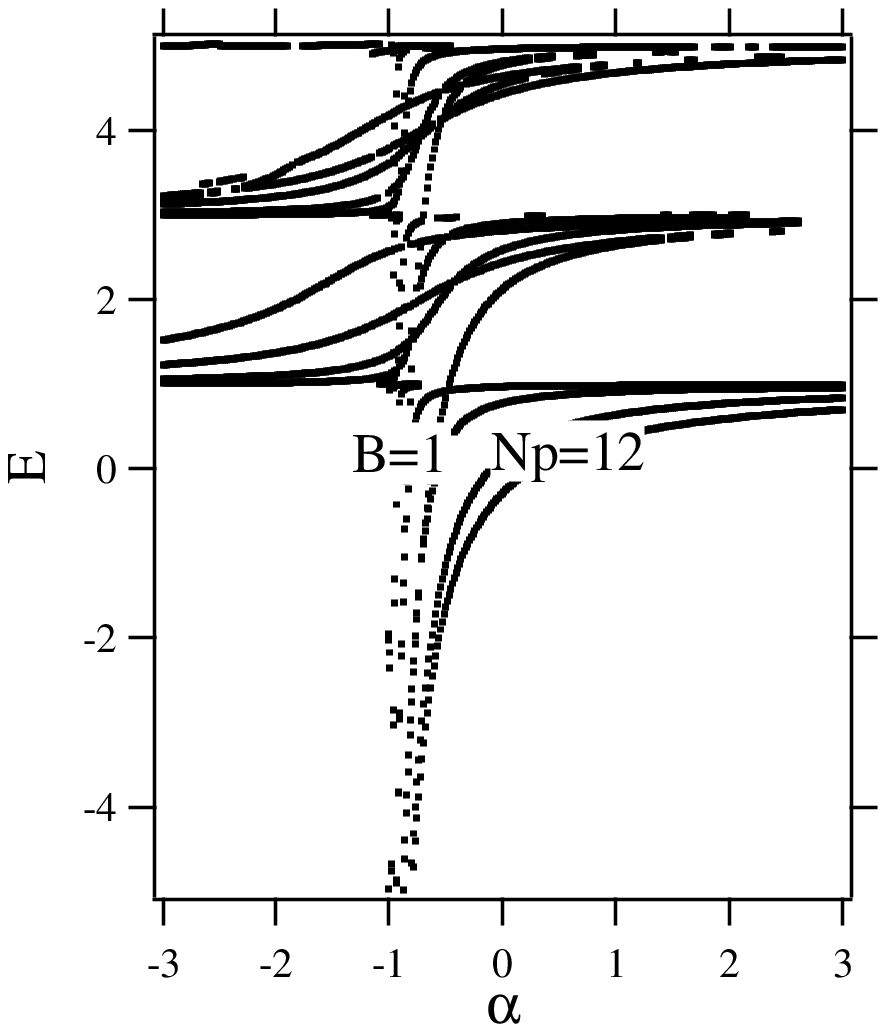}
}} \caption { The energy spectra of a quantum particle  on
2-dimensional plane with $Np=12$ delta functions of strength
parameter $\alpha$ placed equidistantly on a ring of radius $R=1$
and subjected to perpendicular magnetic field of constant strength
$B=1$.  All quantities are in the unit  $\hbar = c = e = 2m* = 1$. }
\end{figure}
%
\begin{figure}
\centerline{
\scalebox{0.45}{
\includegraphics{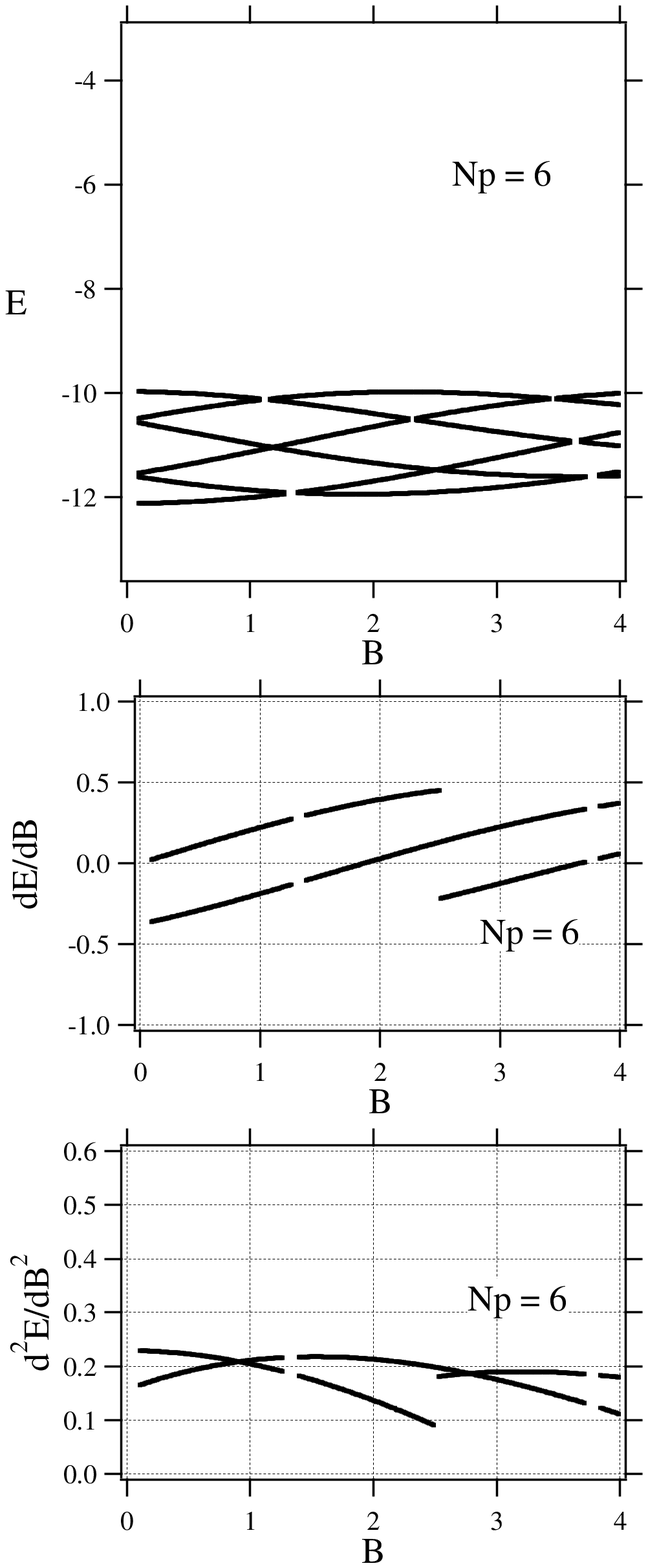}
\includegraphics{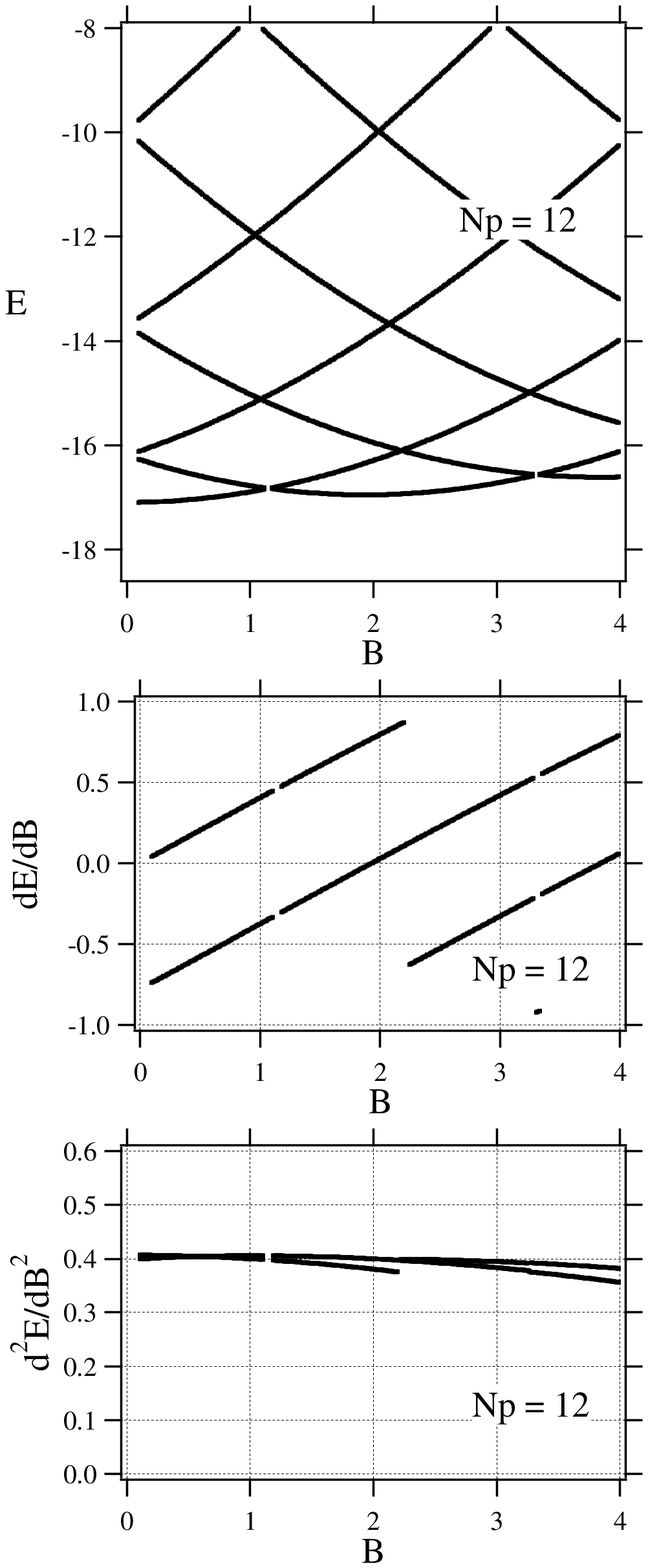}
\includegraphics{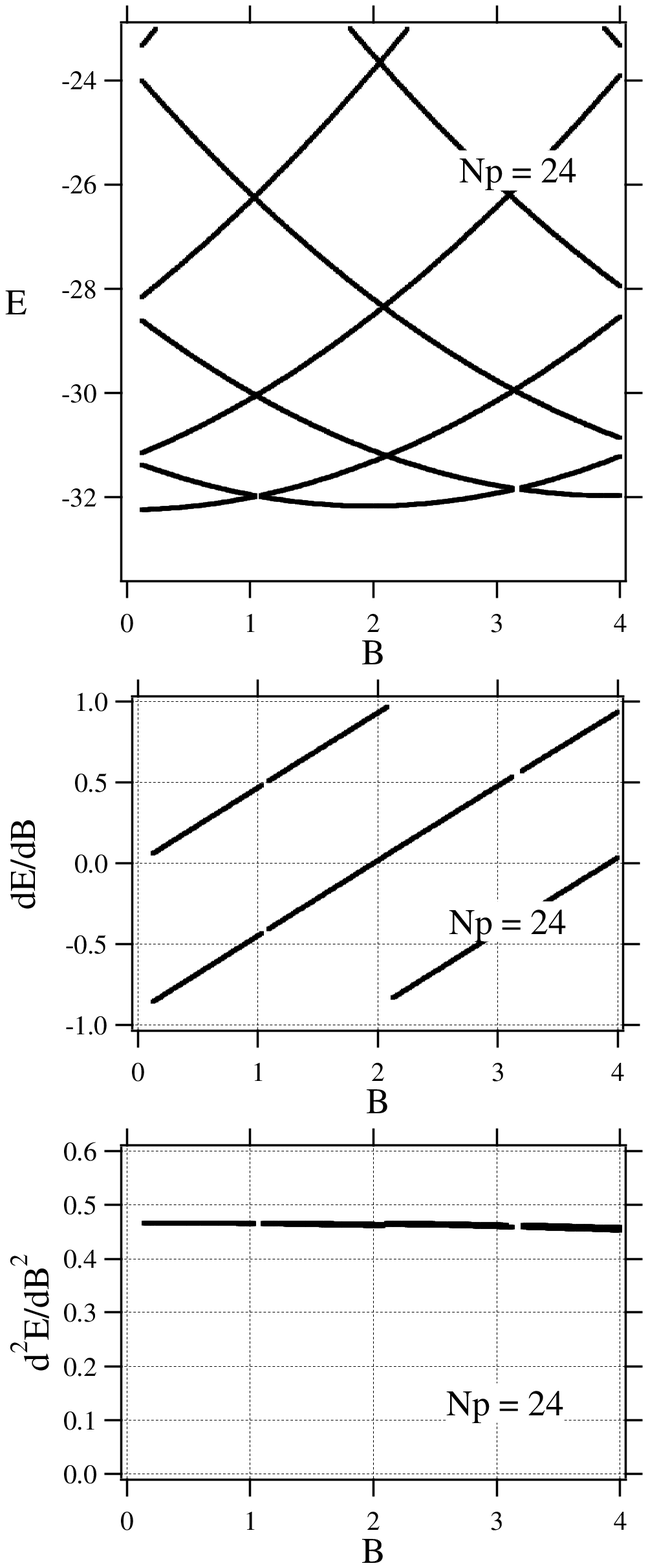}
}} \caption { The energy spectra of a quantum particle on
2-dimensional plane with $Np$ delta function of strength parameter
$\alpha = -1$ placed equidistantly on a ring of radius $R=1$ with
perpendicular magnetic field of constant strength $B$. }
\end{figure}
%
\begin{figure}
\centerline{
\scalebox{0.40}{
\includegraphics{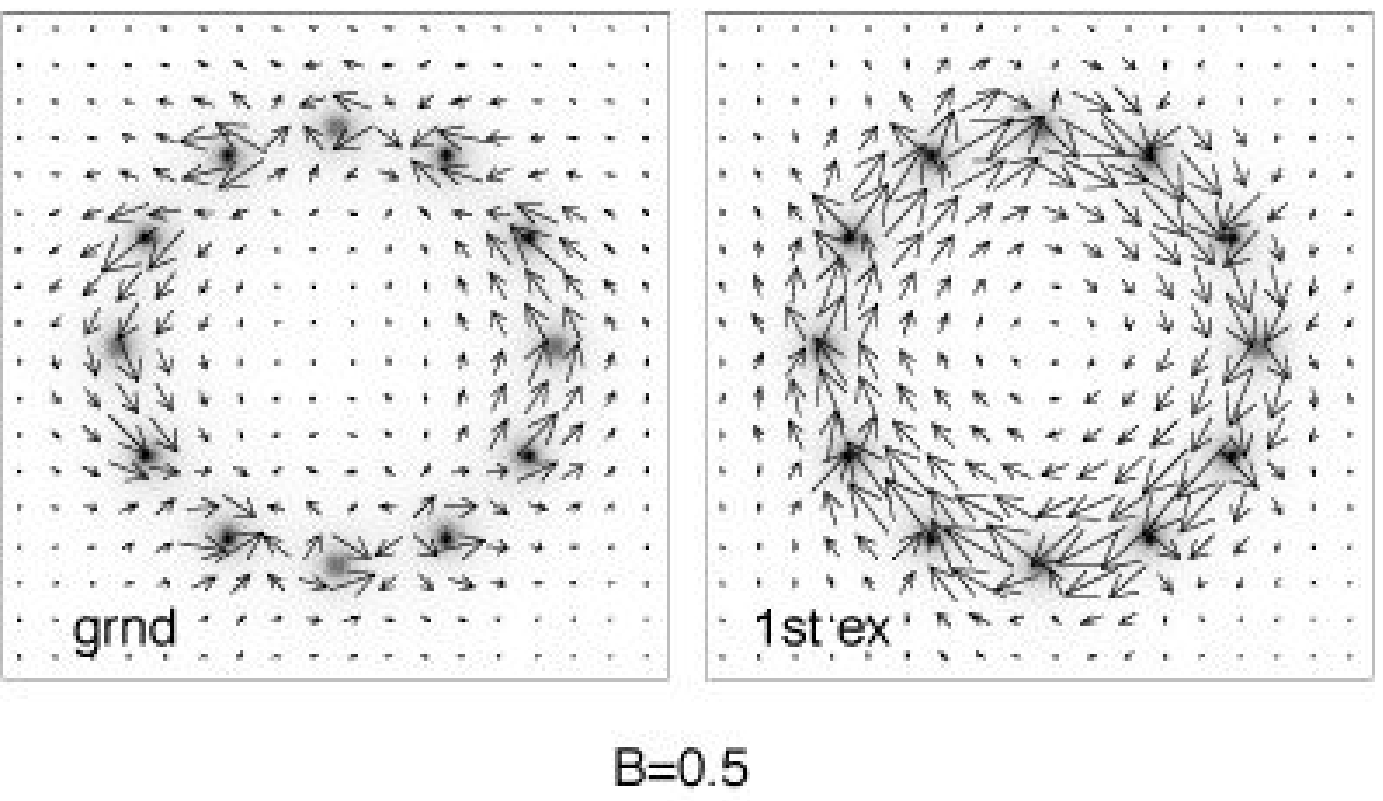}
\includegraphics{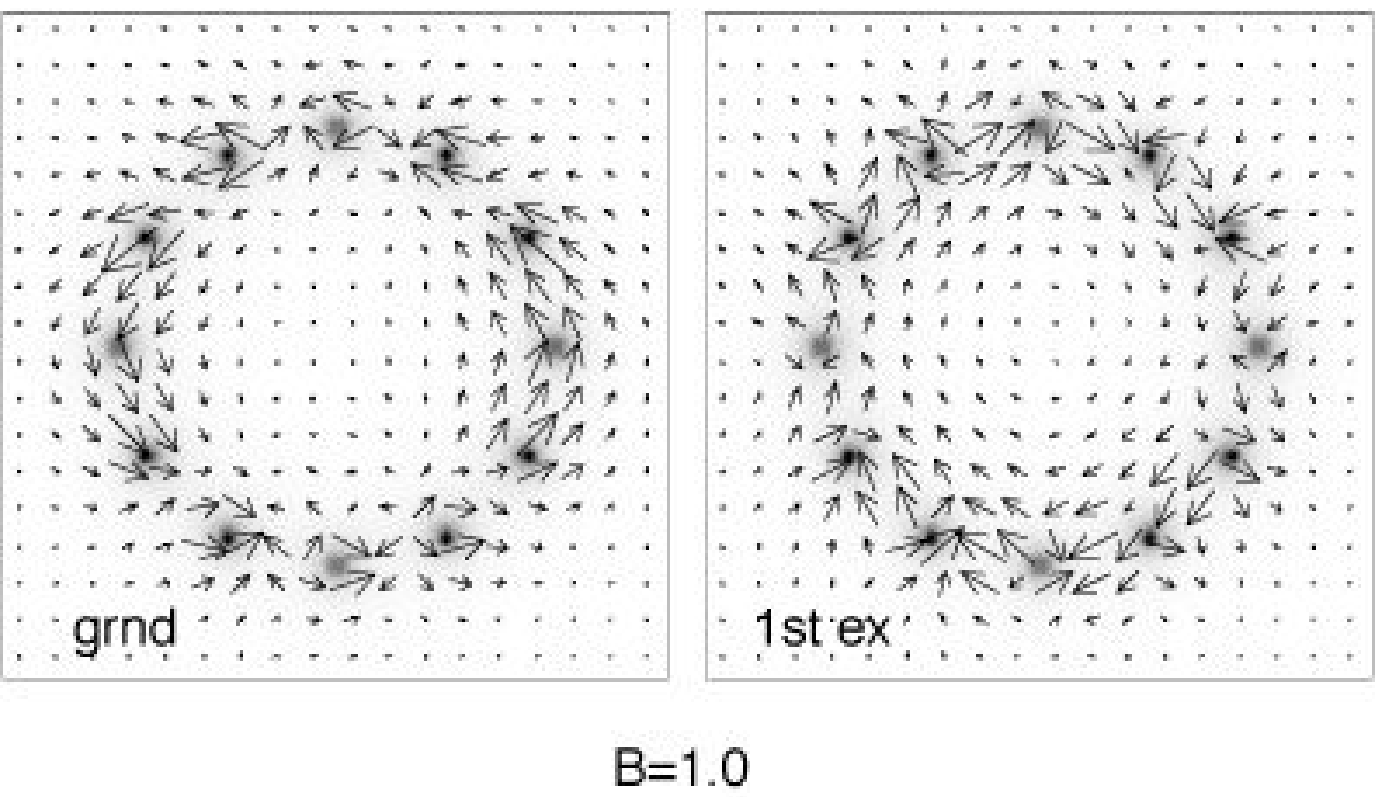}
}}
\centerline{
\scalebox{0.40}{
\includegraphics{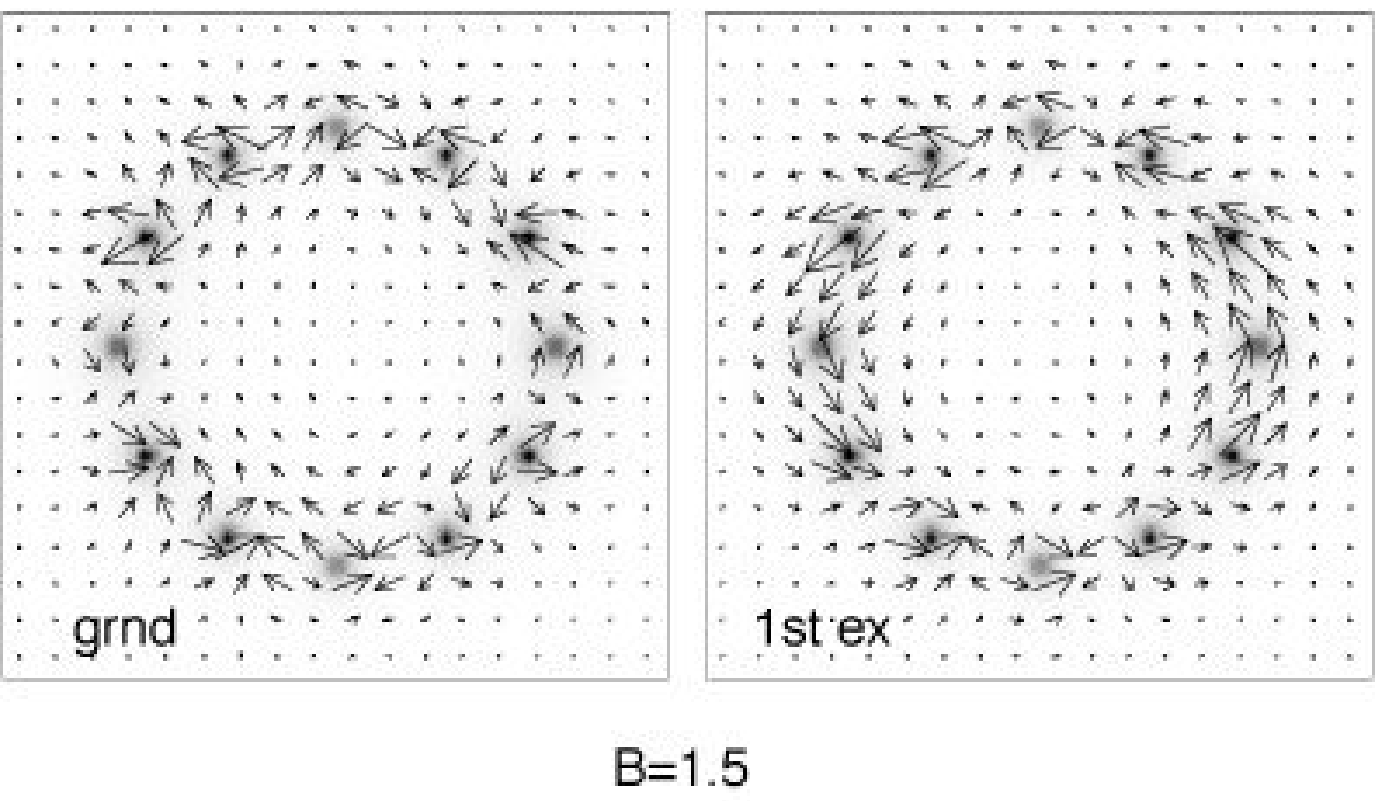}
\includegraphics{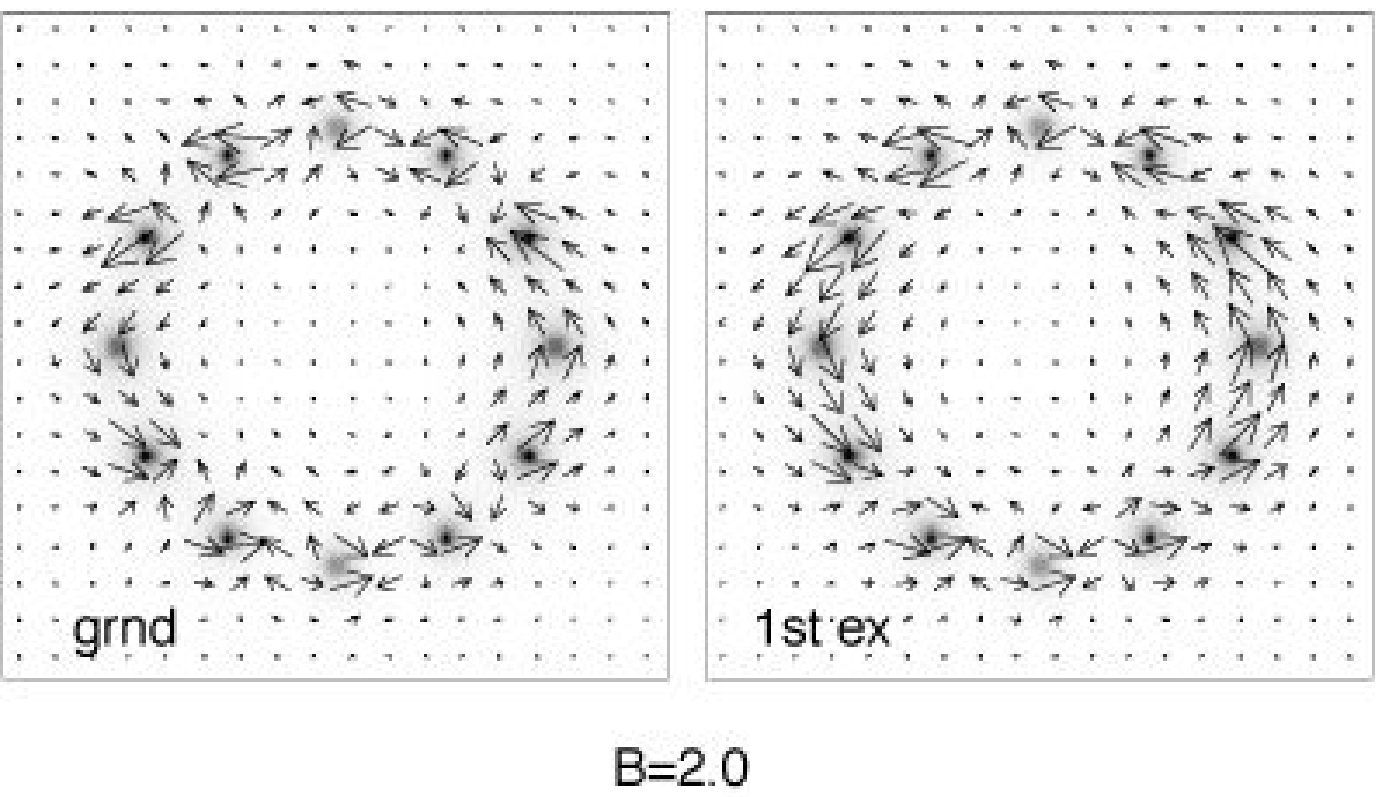}
}}
\caption
{
Profile of probability current for $Np = 12$ delta-functions
of strength parameter $\alpha$ placed equidistantly on a ring of
radius $R=1$ and subjected to perpendicular magnetic field of
constant strength $B=0.5$, $B=1.0$, $B=1.5$ and $B=2.0$.
}
\end{figure}
%
\begin{figure}
\centerline{
\scalebox{0.45}{
\includegraphics{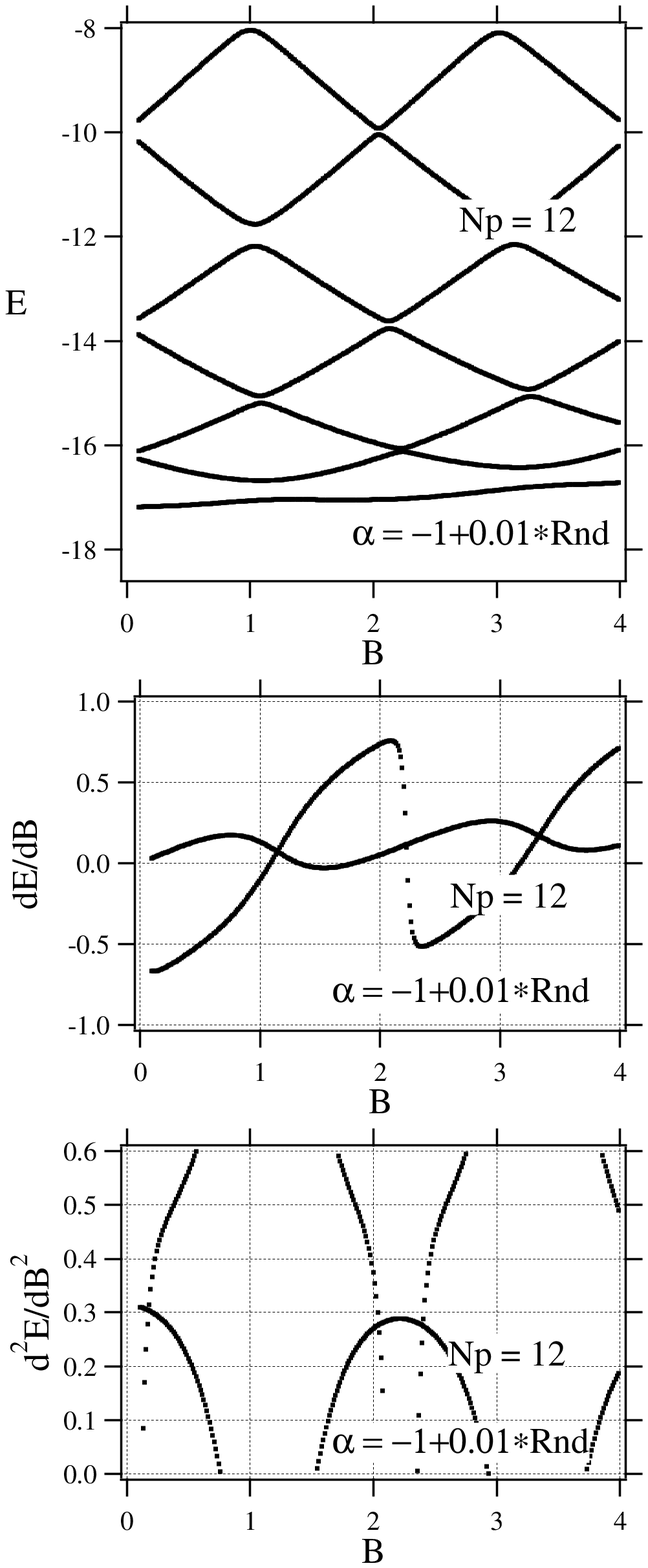}
\includegraphics{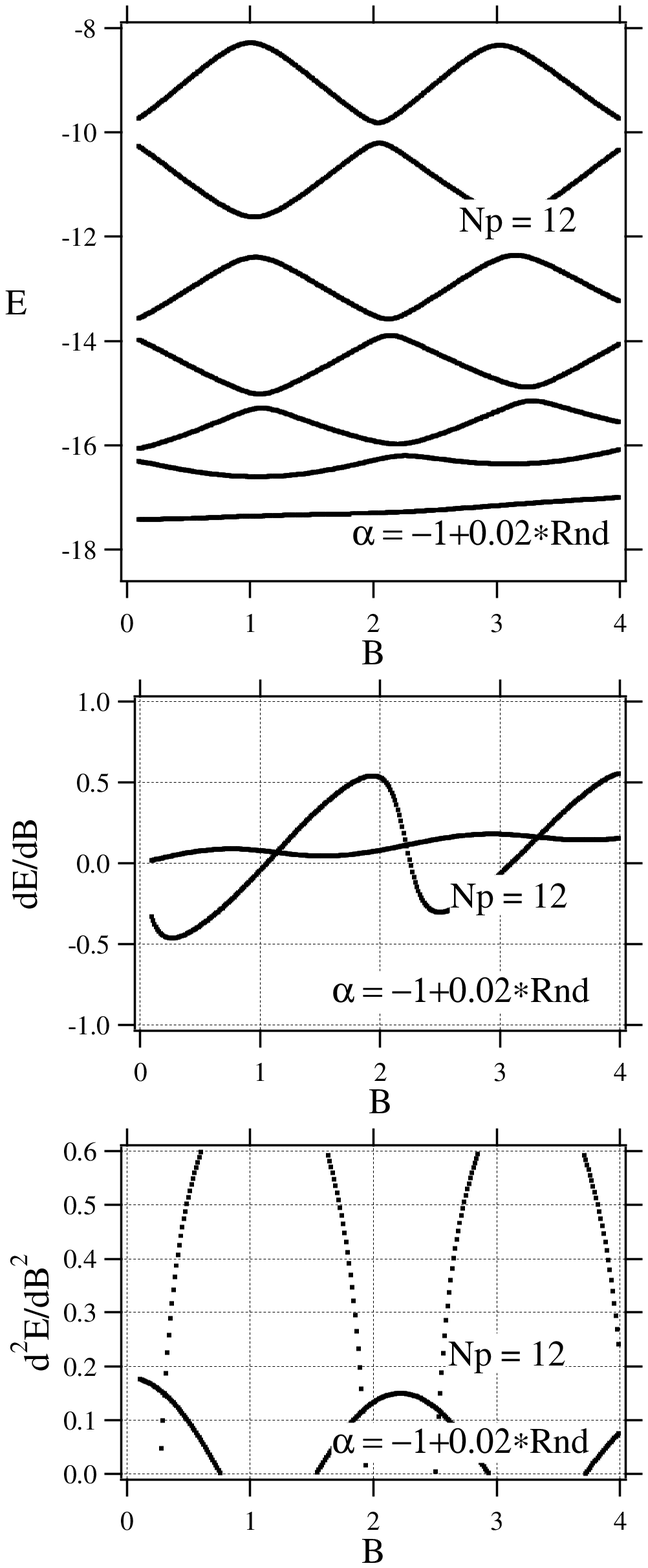}
\includegraphics{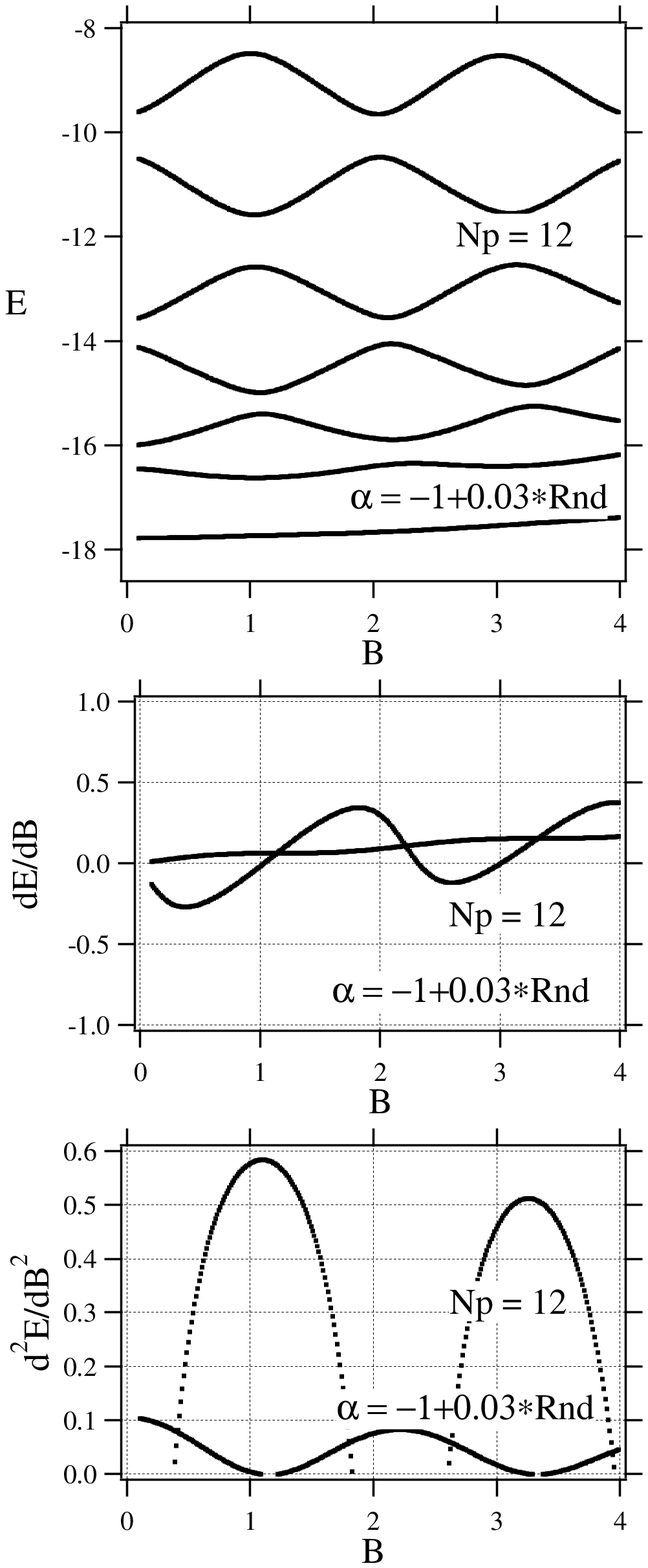}
}} \caption { The energy spectra of a quantum particle on
2-dimensional plane with $Np$ delta function of fluctuating
strength parameter $\alpha=-1+\Delta\alpha$ placed equidistantly
on a ring of radius $R=1$ with perpendicular magnetic field of
constant strength $B$. }
\end{figure}
%
\begin{figure}
\centerline{
\scalebox{0.40}{
\includegraphics{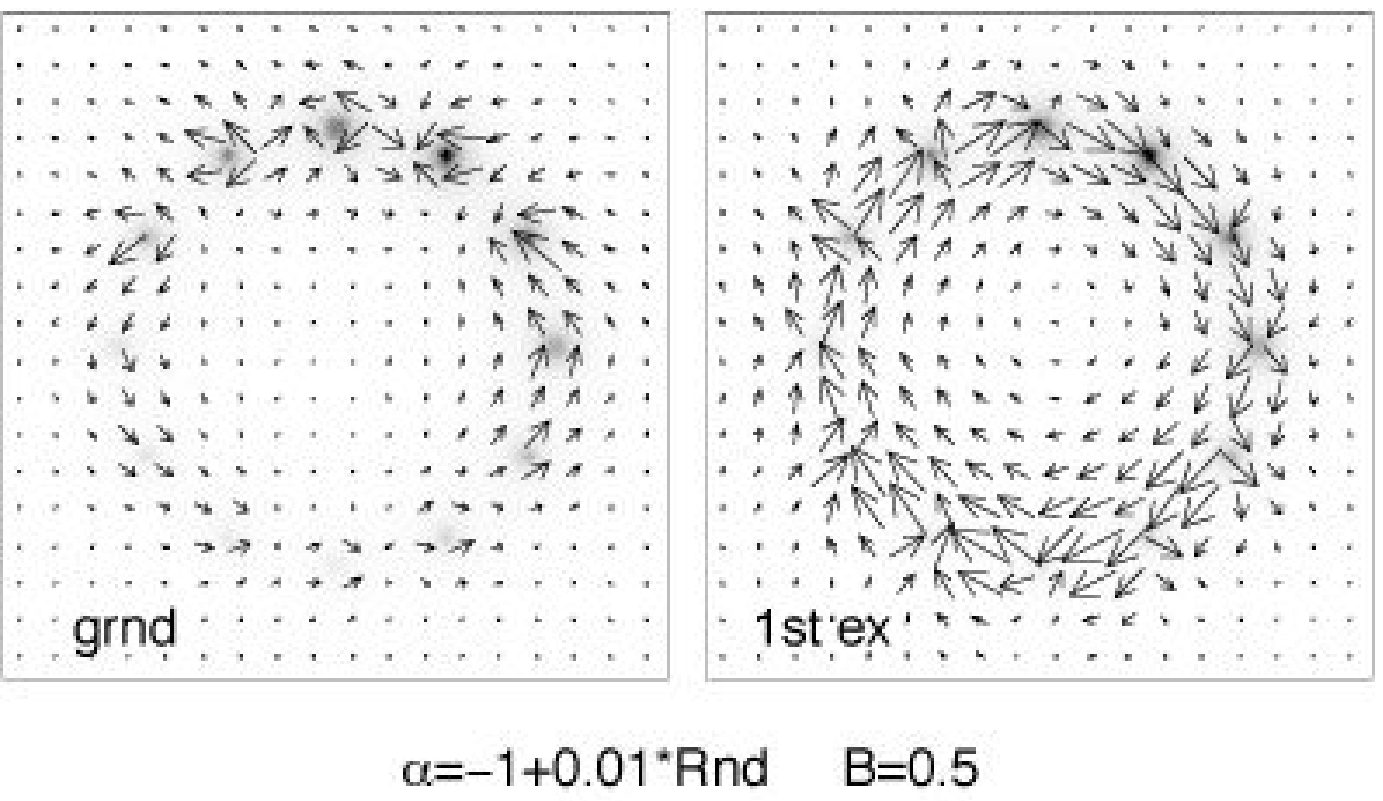}
\includegraphics{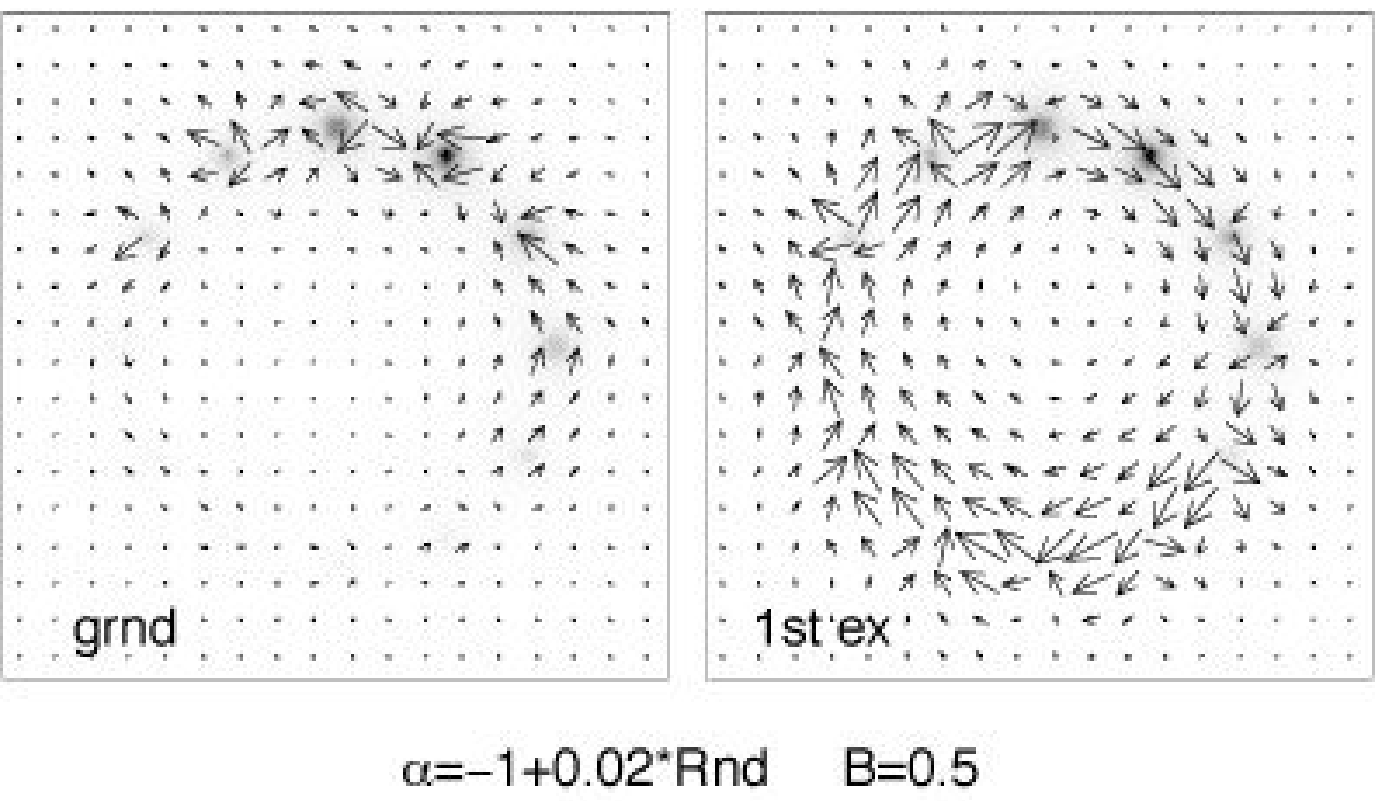}
}}
\centerline{
\scalebox{0.40}{
\includegraphics{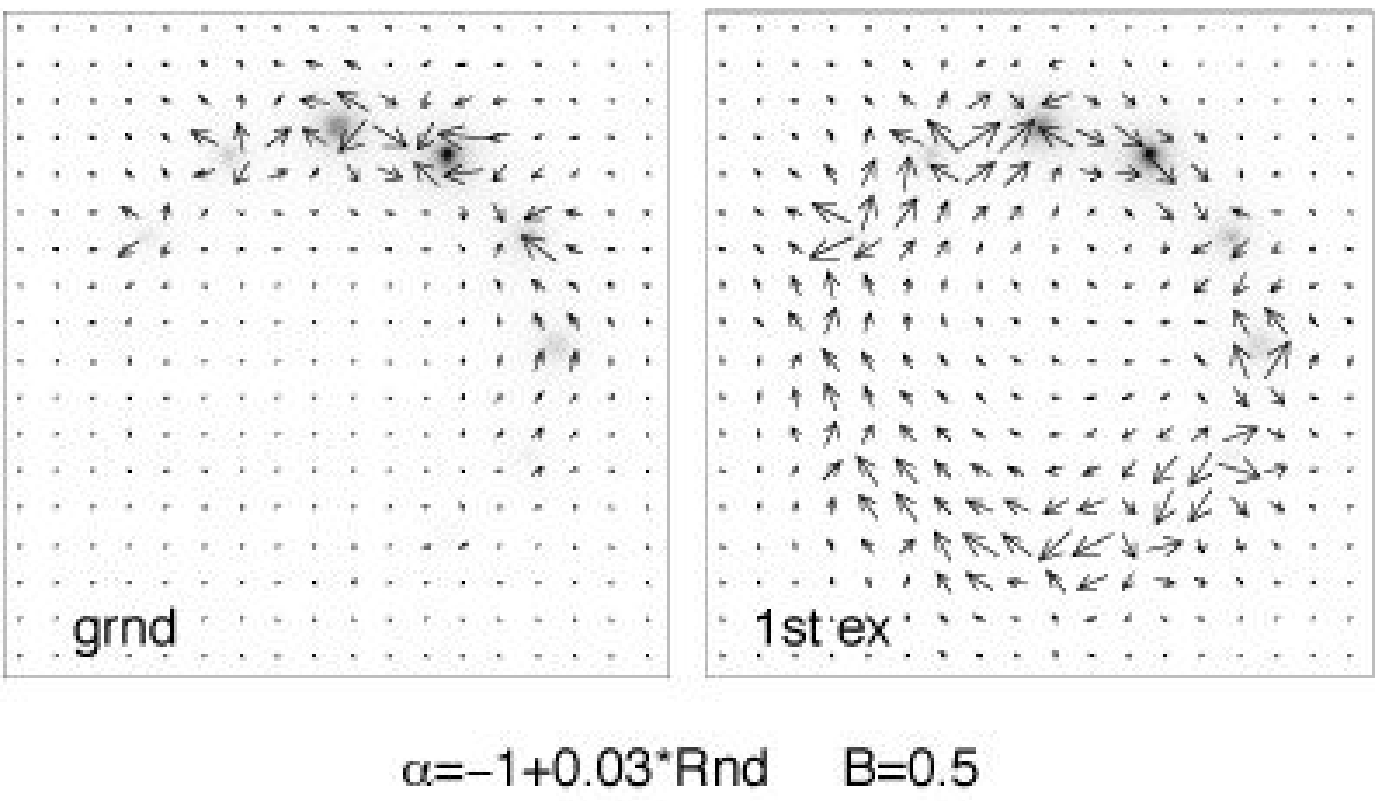}
}}
\caption
{
Profile of probability current for $Np = 12$ delta-functions
of fluctuating strength parameter $\alpha = -1+\Delta\alpha$ placed
equidistantly on a ring of
radius $R=1$ and subjected to perpendicular magnetic field of
constant strength $B=0.5$.
}
\end{figure}

\end{document}